\documentclass[a4paper,11pt]{article}
\usepackage{aaskaiid}
\usepackage{booktabs}
\usepackage{orcidlink}

\title{Probing cosmological phase transitions with SKAO through nano-Hz gravitational-wave portal}
\ShortTitle{Cosmological FOPTs at SKAO}

\author[1]{Roman Pasechnik\,\orcidlink{0000-0003-4231-0149}}

\ShortName{R.~Pasechnik et al.} 
\author[2]{Ant\'onio~P.~Morais\,\orcidlink{0000-0001-7141-8162}}

\affiliation[1]{Department of Physics, Lund University, 221 00 Lund, Sweden}
\emailAdd{roman.pasechnik@fysik.lu.se}
\affiliation[2]{Laborat\'{o}rio de Instrumenta\c{c}\~{a}o e F\'{i}sica Experimental de Part\'{i}culas (LIP), 
Universidade do Minho, 4710-057 Braga, Portugal}

\abstract{The recent evidence for a stochastic gravitational wave (GW) background at nano-Hz frequencies 
from pulsar timing arrays (PTAs) provides a strong science case for exploring cosmological histories 
and phase transitions at the Square Kilometer Array Observatory (SKAO). The latter cosmological events 
are of primary interest for physics of the early Universe associated with out-of-equilibrium and relaxation 
phenomena in the vicinity the quantum chromodynamics (QCD) epoch as well as those originating from dark sectors 
or other phenomena beyond the Standard Model of particle physics. Such a novel GW portal provides a complimentary
access to particle physics phenomena that have so far escaped detection at particle colliders, and also probes
fundamental aspects of out-of-equilibrium dynamics beyond the perturbation theory and their cosmological 
implications. This chapter aims at elucidating most interesting physics cases, theoretical models and approaches, 
as well as cosmological scenarios that are of particular relevance for future GW measurements at SKAO. We elaborate 
on how such measurements can deepen our understanding of particle interactions and violent processes 
in the early Universe, and discuss possible competitive advantages and complementarity of SKAO in this 
research compared to other experiments.}

\begin{document}
\maketitle

\section{Nano-Hz gravitational-wave portal}
\label{sec:intro}

The emerging field of nano-Hz gravitational-wave (GW) astronomy has undergone a decisive step 
from theoretical expectation to strong observational evidence. This is due to the Pulsar Timing 
Arrays (PTAs) providing remarkable observational ``handles'' on the nano-Hz stochastic GW 
background (SGWB) that feeds directly into cosmological inference. A PTA regularly times 
an ensemble of MSPs and fits their pulse times of arrival with a deterministic timing 
model that includes spin, astrometry, 
binary motion and known propagation effects. Any deviations from this model appear as timing 
residuals. A SGWB induces a low-frequency, correlated contribution to these residuals. 
When the background is approximately scale-free, it is customarily described by a characteristic 
strain spectrum
\begin{equation}
  h_c(f) \;=\; A_{\rm GWB}\,\Big(\frac{f}{f_{\rm ref}}\Big)^{\alpha} \,,
  \qquad f_{\rm ref} = 1~{\rm yr}^{-1} \,,
  \label{eq:hcpf}
\end{equation}
or, equivalently, by a power spectral density of the residuals $P(f)\propto f^{-\gamma}$, 
with indices related by
\begin{equation}
  \alpha \;=\; \frac{3-\gamma}{2} \,.
  \label{eq:alpha-gamma}
\end{equation}
The crucial discriminator of a GW origin is not the redness of the spectrum, which can also arise 
from intrinsic pulsar noise, but the spatial correlations between pulsars. In general relativity, 
an isotropic, Gaussian tensor background produces the Hellings–Downs (HD) correlation as a function 
of angular separation $\theta$ between pulsar pairs \citep{HellingsDowns:1983,BurkeSpolaor:2019CQG}. 
PTA analyses therefore proceed in two steps: first they test for the presence of a common-spectrum 
red noise (CURN) term shared by all pulsars, and then they ask whether allowing that common 
process to carry HD-like correlations improves the fit relative to models with 
only monopolar (clock-like) or dipolar (ephemeris-like) spatial patterns.

In 2023, the North American Nanohertz Observatory for Gravitational Waves (NANOGrav) released its 15-year data set \citep{McLaughlin:2013ira} and reported multiple, 
mutually consistent lines of evidence for a spatially correlated stochastic process 
in the timing residuals of 67 millisecond pulsars (MSPs), with correlations consistent with 
the HD curve \citep{HellingsDowns:1983} expected for an isotropic SGWB 
in the nano-Hz band \citep{Agazie:2023ome,NANOGrav:2023hvm} (for an insightful discussion 
of the HD effect, see e.g.~\citep{Jenet:2014bea}). Almost simultaneously, the European Pulsar 
Timing Array (EPTA) combined with the Indian PTA (InPTA) published its second data release,
independently finding a common-spectrum process with HD correlations in a European–Indian 
data set of comparable duration and sensitivity \citep{Antoniadis:2023xep,Antoniadis:2023xlr}. 
In Australia, the Parkes Pulsar Timing Array (PPTA) reported a compatible common process 
and HD spatial correlations in its DR3 analysis of 32 pulsars timed over 18~years
\citep{Reardon:2023gzh,Zic:2023ppta}. In China, the FAST-based Chinese PTA (CPTA) presented 
its first data release, timing 57 MSPs over 41 months and finding a correlated signal with 
a $\sim 4.6\sigma$ preference for the HD pattern \citep{Xu:2023wog}. 

Conceptually, detections first manifest in PTA analyses as a common red process 
across pulsars. Allowing for HD spatial correlations then upgrades this to 
a bona fide GW background detection, with the HD kernel tracing its classic 
quadrupolar form \citep{HellingsDowns:1983}. Each collaboration independently 
finds strong evidence for a CURN with a steep red spectrum, and each reports 
that the evidence increases further when HD correlations are allowed, yielding 
amplitudes at $f=1\,{\rm yr}^{-1}$ of order a few $\times 10^{-15}$. Taken together, 
these results show that Galactic-scale timing baselines now probe, with high 
statistical significance, metric fluctuations with periods of years to decades, 
opening a GW frequency window ranging from a nHz to sub-$\mu$Hz that previously 
was accessible only via indirect cosmological 
bounds \citep{BurkeSpolaor:2019CQG,Taylor:2025SKAReview}.

The regional efforts are being knit together by the International Pulsar Timing Array (IPTA), 
whose second data release demonstrated that the joint analyses improve sensitivity compared 
to that of individual arrays, broaden sky coverage and help control systematics enabling 
the SGWB to be studied as a genuine astrophysical observable \citep{Antoniadis:2022yph}. 
At the same time, differences remain between arrays in the recovered spectral indices and 
in the preference for a single power law versus a broken or running spectrum. Much of this 
dispersion is now understood to arise from analysis choices: the flexibility allowed in 
pulsar-specific red-noise models; how time-variable dispersion measure and Solar-wind 
fluctuations are modeled using multi-frequency or wideband timing; and how Solar-System 
ephemeris and clock errors are treated as array-wide correlated processes
\citep{Antoniadis:2023xlr,Agazie:2023ome,Reardon:2023gzh,Xu:2023wog}. Nevertheless, 
the key message is that the existence of a nano-Hz background with approximately HD-like 
correlations is robust, while the detailed shape of its spectrum and its precise amplitude 
carry analysis-driven uncertainties at the $\mathcal{O}(1\sigma)$ level that must be propagated 
into cosmological parameter inference \citep{BurkeSpolaor:2019CQG,Taylor:2025SKAReview}.

The most economical interpretation of the nano-Hz background remains 
the incoherent superposition of signals from a cosmological population 
of inspiralling supermassive black-hole binaries (SMBHBs) formed during 
galaxy assembly, which generically yields a characteristic strain 
$h_c(f)\propto f^{-2/3}$ at PTA frequencies \citep{Phinney:2001,JaffeBacker:2003,Sesana:2008}. 
This picture is consistent with hierarchical galaxy-formation models and with
electromagnetic searches for dual active galactic nuclei (see \citep{BurkeSpolaor:2019CQG}
for review). However, from the outset it has been recognized that alternative, 
genuinely cosmological origins can reproduce the observed amplitudes and spectral 
slopes once spectral breaks, broadened peaks or multi-component spectra are allowed. 
Systematic studies that jointly fit NANOGrav, EPTA+InPTA, PPTA and CPTA data to SMBHBs,
first-order phase transitions (FOPTs), cosmic (super)strings, scalar-induced GWs, or
mixtures thereof show that several cosmological hypotheses provide fits that are
statistically competitive with -- and in some priors, preferred over -- the purely 
astrophysical model, while degeneracies still remain between spectral index, 
amplitude and foreground composition (for more details, see e.g.~\citep{Goncalves:2025uwh,Figueroa:2024pta,Wu:2023pta,Zu:2024mirror,Fujikura:2023dpt,
Gouttenoire:2023ftt,Madge:2023jhep} and references therein). 

In particular, a strongly supercooled FOPT at temperatures around or below the QCD 
scale yields the SGWB spectrum that naturally peaks today at $f\sim\,$nHz, can match 
PTA-inferred energy densities and can be realized in extended Higgs, composite, or 
nearly conformal models (see \citep{Caprini:2018csg} for a comprehensive review). 
The same band is compatible with stochastic backgrounds from cosmic-string networks. 
Fits of stable local strings to PTA data typically point to tensions in the range 
$G\mu \sim 4\times 10^{-11}$–$10^{-10}$, while cosmic superstrings or networks of 
global strings can reproduce the observed amplitude with somewhat smaller tensions 
$G\mu \sim 10^{-12}$–$10^{-11}$ once realistic loop distributions and intercommutation
probabilities are taken into account \citep{Ellis:2020ena,Blasi:2020mfx,
Ellis:2023tsl,Wang:2023CSloops,Kume:2024CSbounds}. As of now, PTA data 
alone neither exclude the fundamental physics scenarios for SGWB production nor single 
them out uniquely. However, together with cosmological (e.g.~CMB) and astrophysics data 
as well as other New Physics bounds (e.g.~from colliders), they already carve out non-trivial 
regions of parameter space in some of the well-known theoretical models
\citep{Caldwell:2022DEUGW,Bian:2025GWcosmology,Roshan:2025FirstSecond}.

The Square Kilometre Array Observatory (SKAO) will inherit robust PTA analysis pipelines and
will significantly expand the frontiers of GW astrophysics beyond the current state of the art. 
By the time SKA1-MID and, later, SKA2 contribute timing data to the global PTA, a nano-Hz SGWB 
is expected to be firmly established by multiple independent arrays, and the dominant noise 
and systematics channels will have standard treatments in the analysis pipelines
\citep{Antoniadis:2022yph,Taylor:2025SKAReview}. Large collecting area, multi-beam capability, 
and excellent southern-sky coverage of SKA1-MID, and especially SKA2, will enable SKAO to
substantially (by about an order of magnitude) increase the number of PTA-quality MSPs, mainly 
in the southern hemisphere, to improve timing precision for the best sources to the $\lesssim 
100$\,ns regime and angular sampling of the HD curve. It would deliver significantly longer 
effective baselines particularly when SKAO data are combined with observations at NANOGrav, 
EPTA/PPTA, MeerKAT, FAST, Parkes/Murriyang, uGMRT and other facilities in the global IPTA-style 
effort \citep{Janssen:2015SKA,Stappers:2018rsta,Liu:2011cka,Smits:2009,SKAO:Perf2024}. 

Early MeerKAT PTA (MPTA) results have already demonstrated additional evidence for HD-like
correlations in an independent data set, underscoring the value of diverse instruments and sky 
coverage \citep{Miles:2025MPTA}. In the SKAO era, PTAs will not merely reinforce the nano-Hz SGWB
detection already indicated by current arrays, but will turn it into a precision observable. 
Forecasts including SKA-like sensitivities and using fast PTA simulators show that with 
a few hundred well-timed MSPs, a SKA-era PTA will be able to measure the overall amplitude 
and spectral slope of the dominant background with high accuracy and that broken or peaked spectra 
can be distinguished from a simple power law, even in multi-component models
\citep{Babak:2024fastPTA,Lamb:2023Rapid,Gersbach:2025SpatialSpectral}. For finite-duration sources
such as first-order phase transitions, the GW spectrum is expected to exhibit a causal $f^{3}$ rise 
at low frequencies and a turnover near the peak \citep{Caprini:2018csg,Hindmarsh:2017}, features that
become resolvable once SKAO extends PTA baselines and improves sensitivity in the $10^{-9}$–
$10^{-7}\,\mathrm{Hz}$ band. Dedicated studies show that upgraded PTAs can constrain or detect dipole
and quadrupole anisotropies at the percent level and begin mapping the angular structure of the
background \citep{Pol:2022Anisotropy,Depta:2024Anisotropy}. They will also be able to test for
circular polarization and non-tensorial polarization modes, providing probes of parity violation 
and alternative theories of gravity
\citep{SatoPolito:2022CircPol,JimenezCruz:2024CircPol,Lee:2008NonEinsteinian,Chamberlin:2012AltPol,
Cornish:2018AltGravity}. Finally, the combination of SKAO with existing arrays will enable joint
component-separation analyses that disentangle an astrophysical SMBHB foreground from additional
cosmological contributions to the nano-Hz background
\citep{Babak:2024fastPTA,Lamb:2023Rapid,Gersbach:2025SpatialSpectral}. These capabilities will
establish what we refer to as the nano-Hz GW portal: a direct, GW-based probe of the low-temperature
thermal history of the Universe, alternative models of gravity, and hidden particle sectors that are
either secluded from the Standard Model (SM) or too feebly coupled to be accessible at 
colliders or in CMB $B$-mode searches 
\citep{BurkeSpolaor:2019CQG,Taylor:2025SKAReview,Caldwell:2022DEUGW,Bian:2025GWcosmology,
Roshan:2025FirstSecond}.

From the perspective of early-Universe physics, the nano-Hz GW portal is especially
attractive because it singles out precisely those scenarios that are otherwise hardest 
to probe with laboratory experiments or CMB observables. Such scenarios as late or 
low-temperature FOPTs in dark (or mirror) QCD and Yang-Mills sectors, supercooled
transitions in (nearly) conformal sectors, and non-standard expansion histories with 
early matter domination admit parameter ranges in which their GW spectra peak naturally 
in the SKA/PTA band without extreme fine-tuning \citep{Goncalves:2025uwh,Zu:2024mirror,Schwaller:2015PRL,Addazi:2018ctp,Aoki:2017,Morgante:2023} 
(see also \citep{Caprini:2018csg}). In such models, bubble nucleation (their growth
and collisions) and the resulting long-lived acoustic waves and magnetohydrodynamic
turbulence act as efficient GW sources. The SGWB peak frequency and spectral width today
are governed primarily by the transition temperature, the inverse-duration parameter
$\beta/H_*$, and the wall velocity, leading to signals around $10^{-8}$–$10^{-7}\,\mathrm{Hz}$ 
for transitions at or below the QCD scale \citep{Caprini:2018csg,Hindmarsh:2017}. Many of these 
constructions are motivated independently of GWs -- e.g.~by Dark Matter production, the dilution 
of unwanted relics, or the resolution of mild cosmological parameter tensions -- so that SKAO 
observations directly intersect model building in particle cosmology \citep{Caldwell:2022DEUGW,Bian:2025GWcosmology,Roshan:2025FirstSecond}. 

The SKAO band is also complementary to that of higher frequency one at LISA, which is most 
sensitive to FOPTs at substantially higher temperatures, $\gtrsim 10^2\,\mathrm{GeV}$ such 
as electroweak FOPTs \citep{Caprini:2020pt}, and to ground-based interferometers such as 
Advanced LIGO, Virgo and KAGRA (and future facilities like the Einstein Telescope and Cosmic 
Explorer), which probe stochastic backgrounds in the audio band $\sim 10$–$10^3\,\mathrm{Hz}$ 
from much higher-temperature cosmological processes and from unresolved populations of compact
binaries. Combining SKAO with these instruments would chart $\Omega_{\rm GW}(f)$ over many
decades in frequency, following the multi-band programme laid out in \citet{Lasky:2016PRX}.
In the rest of this chapter, we therefore treat a small set of observable quantities -- primarily 
the amplitude and slope of the nano-Hz background, any detected peak or turnover frequency, and, 
at a later stage, its degree of anisotropy -- as the PTA inputs that SKAO will provide for probing
cosmological phase transitions through the nano-Hz GW portal.

\section{Cosmological first-order phase transitions as nano-Hz GW sources}
\label{sec:fopts}

\begin{figure}[t]
\centering
\includegraphics[width=0.49\linewidth]{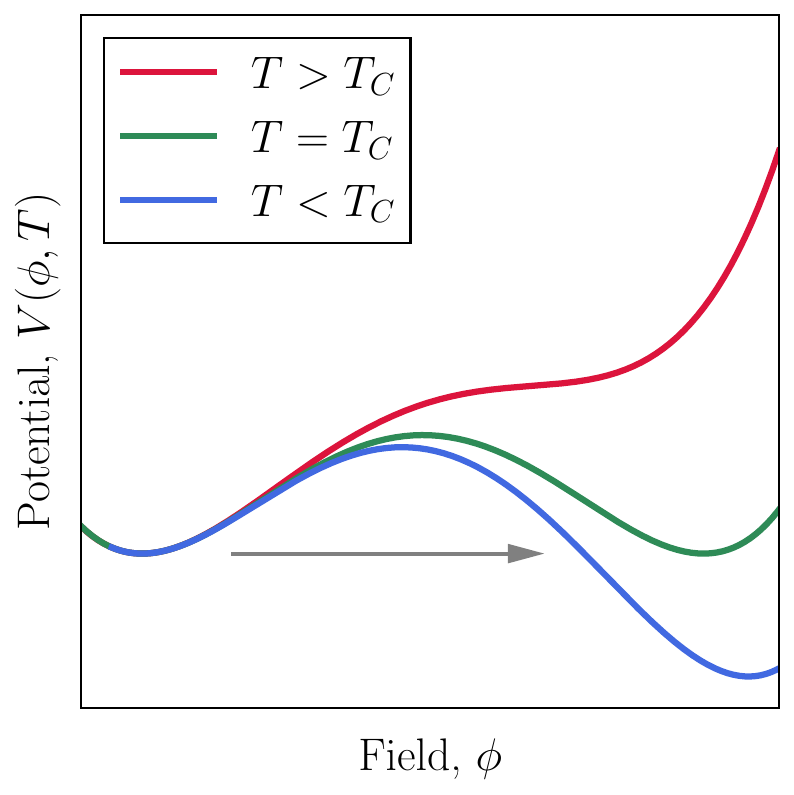}
\includegraphics[width=0.49\linewidth]{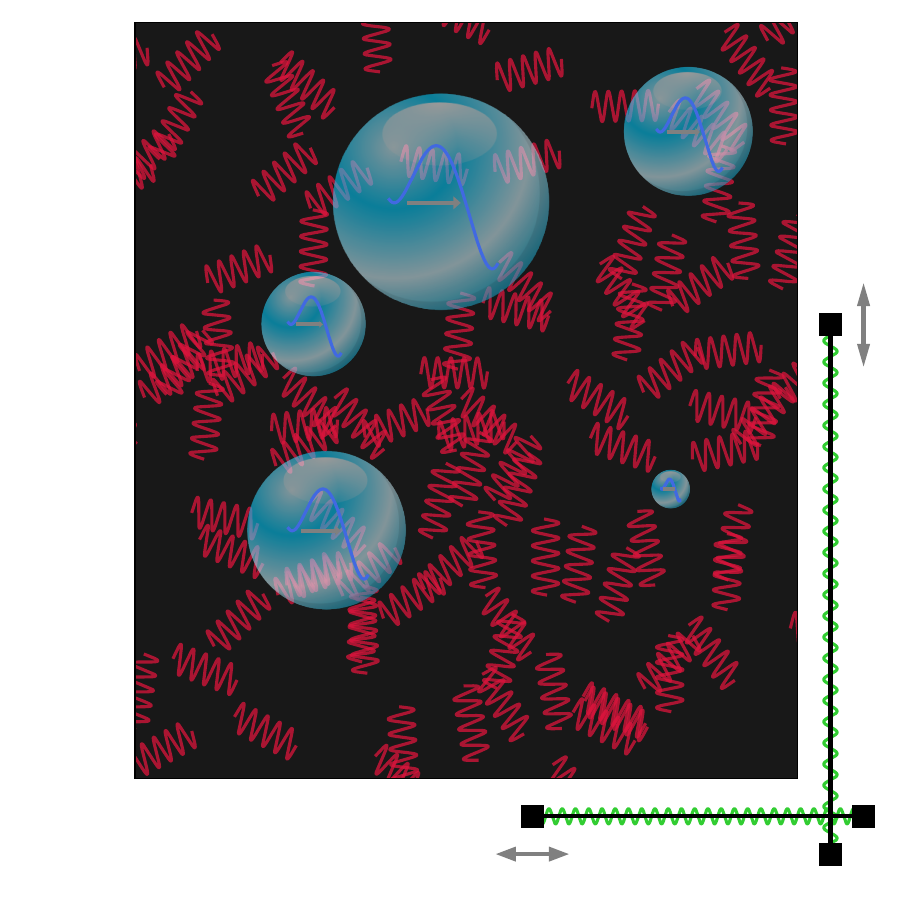}
\caption{Schematic overview of cosmological FOPTs, adapted from Ref.~\citep{Athron:2023ghh}. 
  \emph{Left:} As the Universe cools, the effective potential develops a new 
  minimum away from the origin, separated from the symmetric phase by a barrier, such 
  that the system, previously trapped in a metastable (false) vacuum, undergoes a FOPT 
  to a true vacuum.
  \emph{Right:} The transition proceeds via quantum (or thermal) tunneling: bubbles 
  of the true vacuum nucleate stochastically and expand in the surrounding relativistic plasma, 
  eventually colliding and percolating. This process generates anisotropic stresses that act 
  as a source of stochastic GWs~\citep{Caprini:2015zlo,Caprini:2019egz,
  Hindmarsh:2020hop}.}
  \label{fig:FOPT-overview}
\end{figure}

As discussed above, the nano-Hz SGWB hinted at by current PTA data opens a unique observational 
window to New Physics scenarios and to cosmological histories that are difficult to probe 
by other means. A particularly well-motivated and popular class of scenarios involves cosmological FOPTs, 
in which a system of fields tunnels from a metastable high-temperature (or high-$T$) phase to a stable 
low-$T$ phase via nucleation of vacuum bubbles as illustrated in Fig.~\ref{fig:FOPT-overview}. Indeed, 
as the Universe cools, the finite-$T$ effective potential develops a new minimum separated from 
the symmetric phase by a barrier, so that the transition proceeds via the nucleation and growth 
of bubbles of the true vacuum. The resulting out-of-equilibrium dynamics, associated with bubble 
collisions, the excitation of acoustic waves in the plasma, and the generation of 
magnetohydrodynamic (MHD) turbulence, naturally sources a SGWB in the early Universe 
(for a comprehensive review on this dynamics, see e.g.~Ref.~\citep{Athron:2023ghh}).
Let us start with an overview, and then adapt to the PTA/SKAO regime, the modern formalism 
originally developed in the context of electroweak and high-scale FOPTs~(see e.g.~Refs.~\citep{Quiros:1999jp,Espinosa:2010hh,Caprini:2015zlo,Caprini:2019egz,Hindmarsh:2020hop}).

\subsection{Theoretical description of cosmological FOPTs}
\label{sec:thermo}

The starting point for analyzing a cosmological phase transition is the finite-$T$ effective potential
\begin{equation}
  V_{\rm eff}(\{\phi_i\},T)
  \;=\; V_0(\{\phi_i\}) \,+\, \Delta V_{T=0}(\{\phi_i\}) \,+\, \Delta V_T(\{\phi_i\},T)\,,
\end{equation}
where $V_0$ is the tree-level scalar potential, $\Delta V_{T=0}$ encodes quantum (Coleman-Weinberg) 
corrections, and $\Delta V_T$ the thermal contributions from all fields in the
plasma~\citep{Dolan:1973qd,Coleman:1973jx,Kirzhnits:1976ts,Quiros:1999jp,Kapusta:2006pm}. At high temperatures $T$, 
the thermal corrections generate effective quadratic and cubic terms, which may yield a barrier between the 
minima (phases) as shown schematically in Fig.~\ref{fig:FOPT-overview}. In this case, the transition between 
the competing phases proceeds via thermal tunneling, thus, being of the first-order type.

The free energy difference between the false and true vacua defines the transition strength. For a single 
order parameter $\phi$, and neglecting subdominant temperature dependence of the relativistic degrees of freedom, 
one introduces
\begin{equation}
  \alpha \;\equiv\;
  \frac{\Delta\rho_{\rm vac}(T_*)}{\rho_{\rm rad}(T_*)}
  \simeq
  \frac{1}{\rho_{\rm rad}(T_*)}\Big[V_{\rm eff}(\phi_{\rm false},T_*) - V_{\rm eff}(\phi_{\rm true},T_*)
  - T_*\,\partial_T\!\big(V_{\rm eff}(\phi_{\rm false},T) - V_{\rm eff}(\phi_{\rm true},T)\big)\big|_{T_*}\Big]\,,
\end{equation}
where $\rho_{\rm rad} = (\pi^2/30)\, g_* T_*^4$ is the radiation energy density at the FOPT epoch, and $T_*$ 
is a characteristic temperature around which most of the transition takes place, typically approximated by 
the percolation temperature $T_p$. Weak transitions have $\alpha \ll 1$, while strongly supercooled 
transitions are characterized by a large $\alpha \gg 1$~\citep{Espinosa:2010hh,Ellis:2019oqb,Athron:2022mmm}, 
while the vacuum energy may dominate the energy budget.

Bubble nucleation is usually described semiclassically by the three-dimensional Euclidean action,
\begin{equation}
  S_3(T) \;=\; 4\pi \int_0^\infty r^2 \,{\rm d}r\,
  \bigg[\frac{1}{2}\bigg(\frac{{\rm d}\phi}{{\rm d}r}\bigg)^2
  + V_{\rm eff}(\phi,T) - V_{\rm eff}(\phi_{\rm false},T)\bigg]\,,
\end{equation}
evaluated on the $O(3)$-symmetric bounce solution $\phi_b(r,T)$ that interpolates between true and false
vacua~\citep{Coleman:1977py,Callan:1977pt,Linde:1981zj}. The thermal tunneling rate per unit volume is
\begin{equation}
  \Gamma(T) \;\simeq\;
  T^4 \bigg(\frac{S_3}{2\pi T}\bigg)^{3/2}
  \exp\!\bigg[-\frac{S_3(T)}{T}\bigg]\,.
\end{equation}
As the Universe cools, $S_3(T)/T$ decreases and nucleation becomes efficient when $\Gamma/H^4 \sim \mathcal{O}(1)$. 
Defining a nucleation temperature $T_n$ by the condition that the probability to nucleate at least one bubble 
in a Hubble volume becomes order unity, one typically finds $S_3(T_n)/T_n \sim \mathcal{O}(140)$ for 
electroweak-scale FOPTs in the early Universe~\citep{Linde:1981zj,Quiros:1999jp}.

The characteristic time-scale of the transition is encoded in
\begin{equation}
  \beta \;\equiv\; -\left.\frac{{\rm d}}{{\rm d}t}\Big(\frac{S_3}{T}\Big)\right|_{t_*}
  \;\simeq\; H_*\, T_* \left.\frac{{\rm d}}{{\rm d}T}\Big(\frac{S_3}{T}\Big)\right|_{T_*}\!,
\end{equation}
and the dimensionless ratio $\beta/H_*$ controls the size of the peak in the resulting GW
spectrum~\citep{Turner:1992tz,Kamionkowski:1993fg,Caprini:2019egz}. Here, $H_*$ is the Hubble 
rate around $T_*$. Small $\beta/H_*$ corresponds to a slow, long-lasting transition, 
increasing the correlation length and enhancing the GW signal, whereas large $\beta/H_*$ 
describes a fast transition with suppressed GW amplitude.

Several physically relevant temperatures must be distinguished. While the nucleation temperature $T_n$ 
has been defined above, the critical temperature $T_c$ is determined by the degeneracy of the minima, 
$V_{\rm eff}(\phi_{\rm false},T_c) = V_{\rm eff}(\phi_{\rm true},T_c)$. Nucleation becomes efficient 
at $T_n < T_c$, when the tunneling rate overcomes the Hubble expansion. The transition actually 
completes when a sufficient fraction of the Universe has converted to the true vacuum, and 
the bubble network percolates. A rigorous treatment defines the percolation temperature $T_p$ 
as the point where the true-vacuum fraction reaches $P_{\rm true}(T_p) \simeq 0.3$-$0.5$, 
obtained from
\begin{equation}
  P_{\rm false}(t) \;=\; \exp\!\left[-\frac{4\pi}{3}
  \int_{t_c}^{t} {\rm d}t' \,\Gamma(t')\, a^3(t')\, R^3(t,t')\right]\,,
\end{equation}
with $R(t,t')$ the bubble radius and $a(t)$ the scale factor~\citep{Turner:1992tz,Ellis:2020nnr}.
In strongly supercooled transitions, $T_p$ may be significantly smaller than $T_n$, and the Universe can briefly enter 
a vacuum-dominated phase~\citep{Espinosa:2010hh,Ellis:2019oqb,Athron:2022mmm}. This regime is particularly relevant 
for nano-Hz GW signals, because the redshifted peak frequency scales roughly as $f_{\rm peak} \propto (T_*/1\,{\rm MeV})$ 
for fixed $\beta/H_*$, so that MeV/GeV-scale transitions naturally populate the PTA band.

The expansion of the bubbles is governed by microphysics at the bubble wall. A competition between the driving 
force (the pressure difference across the wall) and friction from interactions with the plasma leads to a terminal 
wall velocity $v_w$~\citep{Moore:1995si,Konstandin:2014zta}. If friction is strong, the walls reach a subsonic 
or mildly supersonic steady velocity and most of the released vacuum energy is converted into bulk fluid motion. 
In some corners of parameter space the walls can `run away', accelerating towards the speed of light and 
storing a substantial fraction of the energy in the scalar field configuration itself~\citep{Bodeker:2009qy}.
Whether this happens depends on the structure of the potential, the couplings to plasma constituents and the presence of 
light degrees of freedom in the broken phase. For the majority of models relevant to nano-Hz GW sources the walls are 
not fully runaway, and the dominant GW source is the bulk plasma rather than the wall.

Gauge dependence, infrared divergences and resummation schemes can significantly affect the shape of $V_{\rm eff}$ 
and the derived macroscopic parameters~\citep{Patel:2011th,Garny:2012cg,Ekstedt:2022bff}. A consistent approach 
is to combine dimensional reduction to a three-dimensional effective theory with non-perturbative lattice 
simulations, whenever computationally feasible~\citep{Kajantie:1995dw,Kajantie:1996mn,Laine:1998qk,
Gould:2021oba}. For more general BSM models, one typically relies on resummed perturbation theory complemented by dedicated tools such as \texttt{CosmoTransitions}~\citep{Wainwright:2011qy}, \texttt{PhaseTracer}~\citep{Athron:2019nbd}, \texttt{BSMPT}~\citep{Basler:2018cwe} and \texttt{TransitionSolver}~\citep{Athron:2022mmm}, which implement 
improved treatments of tunneling, percolation and transition completion. The review of~\citep{Athron:2023uxb} 
provides a comprehensive overview of these developments and their implications for GW physics.

\subsection{Gravitational-wave production from FOPTs}
\label{sec:gw-prod}

Once bubbles nucleate, their dynamics sources GWs through several channels. At a microscopic level, 
time-varying quadrupole moments associated with the scalar field configuration and the coupled plasma 
generate tensor perturbations. At a macroscopic level, one typically distinguishes three components: 
scalar-field gradients localized near the bubble walls and their collisions, long-lived acoustic waves 
(sound shells) in the plasma, and MHD turbulence generated as the flow becomes non-linear~\citep{Turner:1992tz,Kamionkowski:1993fg,Huber:2008hg,Caprini:2009dy,Hindmarsh:2015qta,
Hindmarsh:2017gnf}.

Early studies focused on the scalar-field contribution using the envelope approximation, which assumes 
infinitesimally thin shells that cease to radiate after collision~\citep{Turner:1992tz,Huber:2008hg}. 
Numerical simulations subsequently showed that, when bubbles propagate in a relativistic fluid, the dominant 
source is the kinetic energy stored in sound waves of the plasma, which can persist for a sizable fraction 
of a Hubble time~\citep{Hindmarsh:2015qta,Hindmarsh:2017gnf}. A smaller but potentially relevant contribution 
arises from fully developed MHD turbulence~\citep{Caprini:2009dy,Caprini:2019egz}.
\begin{figure}[h!]
\centering
\includegraphics[width=0.95\linewidth]{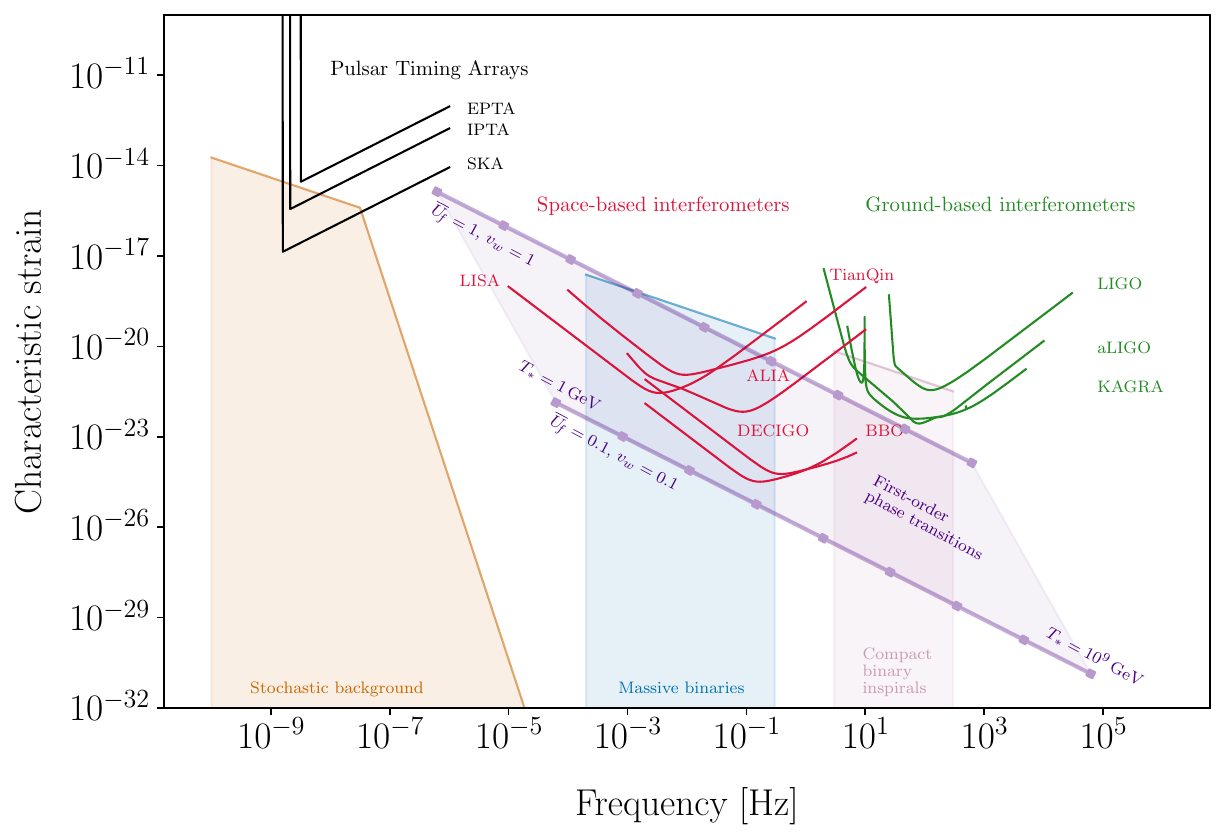}
\caption{Sensitivity curves of the representative current and future GW experiments in the frequency–strain plane, 
together with the characteristic peak frequencies and amplitudes of the sound-wave GW component from a FOPT 
as the transition temperature varies between $T_* = 1$ GeV and $10^9$ GeV, for two benchmarks of the enthalpy-weighted mean
square fluid four-velocity $\overline{U}_{\rm f}^2$ \citep{Hindmarsh:2019phv} and the bubble-wall velocity $v_w$ (bottom line: $\overline{U}_{\rm f}=0.1$, $v_w=0.1$; 
top line: $\overline{U}_{\rm f}=1$, $v_w=1$). The figure is adapted from Ref.~\citep{Athron:2023ghh}.}
  \label{fig:GW_bands-overview}
\end{figure}

For phenomenological applications, the resulting GW spectrum is typically modeled as the sum of three template 
components, each characterized by a peak frequency $f_{{\rm pk},i}$ and a peak amplitude 
$\Omega_{{\rm GW},i}(f_{{\rm pk},i})$ that depend on the FOPT parameters $(\alpha,\beta/H_*,v_w,T_*)$ and 
on efficiency factors $\kappa_i(\alpha,v_w)$ describing the conversion of vacuum energy into each
channel~\citep{Caprini:2015zlo,Caprini:2019egz,Schmitz:2020syl,Hindmarsh:2020hop,Athron:2023uxb}. 
For the sound-wave contribution, which is usually dominant in non-runaway scenarios, one can write 
schematically
\begin{equation}
  \Omega_{\rm sw}(f) h^2 \;\simeq\;
  2.6\times10^{-6}\,
  \left(\frac{H_*}{\beta}\right)\,
  \left(\frac{\kappa_{\rm sw}\alpha}{1+\alpha}\right)^2\,
  \left(\frac{100}{g_*}\right)^{1/3}
  v_w\,
  S_{\rm sw}\!\left(\frac{f}{f_{\rm sw}}\right)\,,
\end{equation}
where $S_{\rm sw}(x)$ encodes the spectral shape, rising as $x^3$ at low frequency and decaying as 
$x^{-4}$ at high frequency, and $f_{\rm sw}$ is the redshifted peak frequency,
\begin{equation}
  f_{\rm sw} \;\simeq\; 1.9\times10^{-5}\;
  \frac{1}{v_w}\,
  \frac{\beta}{H_*}\,
  \left(\frac{T_*}{100~{\rm GeV}}\right)
  \left(\frac{g_*}{100}\right)^{1/6} {\rm Hz}\,.
\end{equation}
For MeV/GeV-scale transitions with $\beta/H_* \sim 10$--$10^3$ this expression places the peak in the PTA band, 
$f\sim 10^{-9}$--$10^{-7}$~Hz, while TeV-scale transitions produce mHz signals in the LISA band as shown 
in Fig.~\ref{fig:GW_bands-overview}.

The scalar-field and turbulence components are described by analogous expressions with different spectral 
shapes and normalization factors. For non-runaway walls the scalar-field contribution is usually
subdominant~\citep{Huber:2008hg,Bodeker:2009qy}, whereas the turbulence component 
inherits a fraction of the kinetic energy of the plasma and contributes a broader, slightly lower-frequency
tail~\citep{Caprini:2009dy,Caprini:2019egz}. The relative 
importance of these channels depends sensitively on the wall dynamics, the efficiency with which 
kinetic energy is converted into turbulence and the lifetime of the acoustic
phase~\citep{Ellis:2020awk,Schmitz:2020syl}.

A central theoretical uncertainty concerns the duration of the sound-wave source. Idealized simulations 
assumed that sound shells persist for a Hubble time, leading to large GW amplitudes. More recent work 
argues that shock formation and non-linear effects terminate the acoustic phase earlier, introducing 
a suppression factor proportional to the ratio of the source lifetime to the Hubble 
time~\citep{Ellis:2020awk,Schmitz:2020syl,Hindmarsh:2020hop}. This can reduce $\Omega_{\rm sw}$ 
by an order of magnitude or more in some regimes. The onset and efficiency of turbulence are also 
not yet fully settled, with different simulations suggesting somewhat different scaling laws and spectral
indices~\citep{Caprini:2009dy,Auclair:2022GWPT}.

On top of these hydrodynamic uncertainties come those associated with the underlying microphysics. 
The mapping from Lagrangian parameters to $(\alpha,\beta/H_*,v_w)$ is affected by gauge dependence, 
truncation of the loop expansion and the treatment of infrared bosonic modes~\citep{Patel:2011th,
Garny:2012cg,Ekstedt:2022bff}. Dimensional reduction combined with lattice simulations provides 
the cleanest framework in theories where it is available, such as SM-like electroweak
sectors~\citep{Kajantie:1996mn,Laine:1998qk,Gould:2021oba}. 
For more complicated BSM sectors, the community has converged on a set of best practices 
and tools, including \texttt{CosmoTransitions}, \texttt{PhaseTracer}, \texttt{BSMPT} 
and \texttt{TransitionSolver}, which implement improved treatments of tunneling, 
percolation and transition completion and allow statistical scans over model parameter
space~\citep{Wainwright:2011qy,Athron:2019nbd,Basler:2018cwe,Athron:2022mmm,Ekstedt:2022bff,Athron:2023uxb}.

For the purposes of SKAO and PTA science, these advances have two key implications. First, they 
enable reliable predictions for the shape and normalization of nano-Hz GW spectra in a wide variety 
of models, including the supercooled conformal and confining dark sectors that we discuss 
in the remainder of this section. Second, they clarify the theoretical error bars on those 
predictions, allowing us to assess how well nano-Hz GW observations -- especially when combined 
with LISA and ground-based GW detectors, CMB and large-scale structure constraints as well as 
particle physics searches -- can discriminate between different FOPT scenarios and 
between FOPTs and alternative nano-Hz GW sources such as astrophysical binaries, 
cosmic strings or primordial magnetic fields.

\subsection{Benchmark FOPT scenarios in particle physics}
\label{sec:benchmarks}

The formalism reviewed above can be realized in a wide variety of extensions of the SM. In order 
to connect cosmological FOPTs to specific nano-Hz GW signatures, it is useful to identify representative 
benchmark classes that capture the typical ranges of $(\alpha,\beta/H_*,v_w,T_*)$ and the resulting spectral 
shapes. In this subsection, we briefly review two such major classes: supercooled dark sectors with MeV/GeV-scale 
symmetry breaking and strongly coupled confining dark sectors that can also furnish composite dark matter.

\subsubsection{Supercooled dark-sector FOPTs}
\label{sec:conformal-dark}

A particularly appealing class of nano-Hz GW sources is provided by conformal dark sectors in which a new scalar field 
$\phi$ acquires a vacuum expectation value through radiative symmetry breaking. The dark sector may contain gauge bosons 
and fermions, and in many constructions is responsible for generating neutrino masses, dark matter or both. The key feature 
is that the same conformal dynamics that generates the dark scale also gives rise to a strongly supercooled FOPT: at high
temperature the plasma is trapped in a metastable symmetric phase, while at some temperature $T_n \ll v_\phi$ bubbles 
of the true vacuum nucleate and eventually percolate, often after a period of vacuum domination~\citep{Ellis:2019oqb,Ellis:2020awk,Athron:2022mmm,Athron:2023uxb}. The thermal parameters in such models can be 
computed using the finite-$T$ effective potential techniques outlined in Section~\ref{sec:thermo}, supplemented by 
numerical tools to solve for the bounce and the percolation history. In many conformal dark sectors one finds 
$\alpha \gg 1$ and $\beta/H_*$ as low as a few tens, signaling a slow, violent transition with strong vacuum domination~\citep{Prokopec:2018tnq,Ellis:2019oqb,Ellis:2020nnr}.

Concrete realizations include conformal neutrino-mass U(1)$'$ models in which the dark scalar couples to right-handed 
neutrinos, generating Majorana masses and enabling low-scale seesaw mechanisms~\citep{Goncalves:2025Supercooled}. The conformal 
potential is then stabilized by additional scalar or fermionic degrees of freedom, and portal couplings to the 
SM Higgs allow energy exchange between the sectors. The same parameters that control neutrino masses (and/or dark matter 
abundance) then fix $(\alpha,\beta/H_*,T_*)$, tying the nano-Hz GW signal to independently testable aspects of particle 
physics. Systematic studies of these models show that a significant region of parameter space can explain the NANOGrav 
signal while remaining consistent with BBN, CMB and laboratory constraints, making predictions for future GW measurements.
Specifically, the typical GW peak frequency scales as $f_{\rm peak}\propto (\beta/H_*)\,T_*$, so for 
$T_*\sim{\rm MeV}$-${\rm GeV}$ and $\beta/H_*\sim 10$-$10^3$ the signal naturally populates the nano-Hz range, 
with amplitudes that can match or exceed current PTA sensitivities~\citep{Goncalves:2025Supercooled}.

\begin{figure}[t]
\centering
\includegraphics[width=0.85\textwidth]{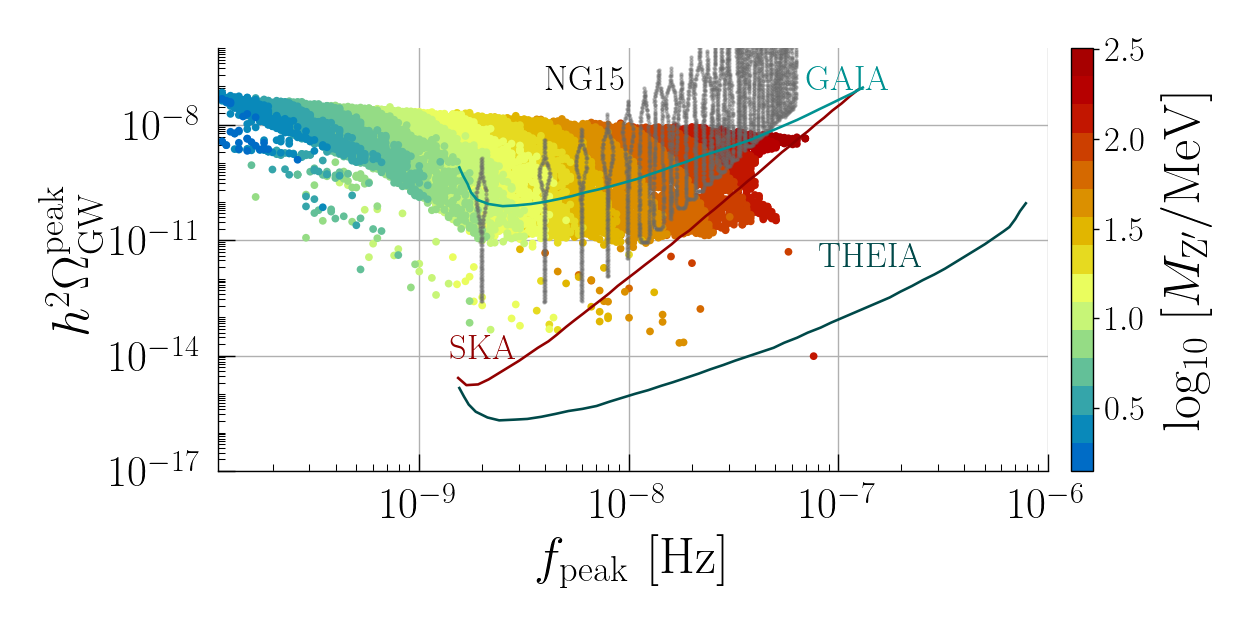}
\includegraphics[width=0.85\textwidth]{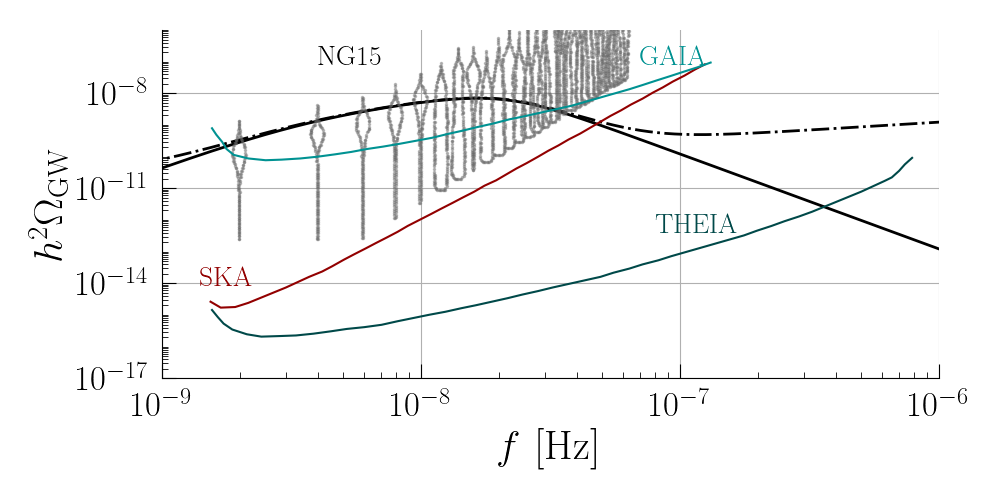}
  \caption{
    Nano-Hz GW signal from a supercooled MeV-scale U(1)$'$ dark-sector phase transition, adapted from Ref.~\citep{Goncalves:2025Supercooled}. 
    \emph{Top:} 
    Scatter plot of the SGWB peak amplitude $h^2\Omega_{\rm GW}^{\rm peak}$ versus peak frequency $f_{\rm peak}$ 
    obtained from a scan over the model parameters, together with the NANOGrav 15-year data 
    (labeled ``NG15''), while the coloured curves indicate the projected sensitivities of GAIA, SKA and 
    THEIA~\citep{Garcia-Bellido:2021zgu,Weltman:2020gw}. 
    \emph{Bottom:} 
    SGWB spectrum $\Omega_{\rm GW}(f)$ for the best-fit benchmark point in Table~I of~\citep{Goncalves:2025Supercooled}. The solid black curve denotes the contribution from 
    the FOPT alone, while the dot-dashed curve includes a SMBHB contribution modeled as a power law.}
  \label{fig:U1p-GW}
\end{figure}

The outcome of the parameter scan and its connection to PTA observables is illustrated in Fig.~\ref{fig:U1p-GW}. 
The top panel shows how the peak frequency and peak amplitude of the SGWB populate 
the $(f_{\rm peak},h^2\Omega_{\rm GW}^{\rm peak})$ plane when the model parameters are varied, with the $Z'$ mass 
in the color scale. The NANOGrav 15-year reconstruction and the projected sensitivities of SKA, GAIA and THEIA 
are overlaid, highlighting a band of MeV-scale supercooled phase transitions that naturally fall in the PTA window. 
The bottom panel displays the full GW spectrum for the best-fit benchmark point in Table~I of~\citep{Goncalves:2025Supercooled},
demonstrating that the FOPT alone can account for the NANOGrav signal, with the SMBHB contribution being subdominant 
across the nano-Hz band.

\begin{figure}[t]
  \centering
  \includegraphics[width=0.9\textwidth]{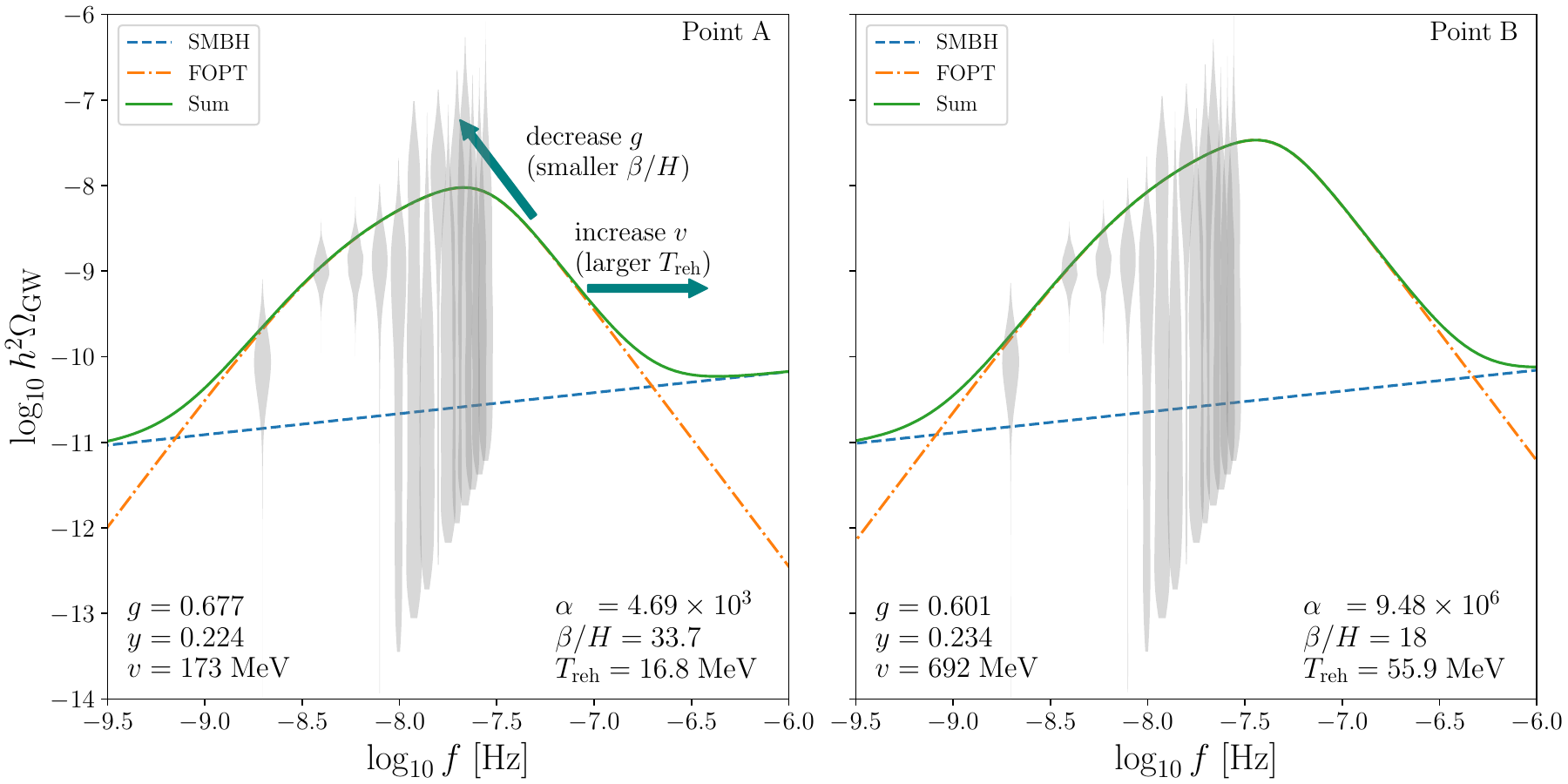}
  \caption{
    GW spectra for two benchmark points in the classically conformal U(1)$'$ dark-sector model with sub-GeV dark matter 
    against NANOGrav 15-year data; adapted from \citep{Balan:2025SubGeVDM}. Curves indicate the contributions from 
    the dark-sector FOPT (dash-dotted lines), a SMBHB background (dashed lines), and their sum (solid lines). 
    \emph{Left:} Best-fit benchmark A in the ``coupled'' dark-sector scenario. 
    \emph{Right:} Best-fit benchmark B in the ``secluded'' dark-sector case, which has a larger symmetry-breaking 
    scale and reheating temperature. 
  }
  \label{fig:Balan-PTA}
\end{figure}

A related but phenomenologically distinct realization of a nano-Hz FOPT in a classically conformal U(1)$'$ sector has 
been proposed in Ref.~\citep{Balan:2025SubGeVDM}. In that setup the dark sector contains a complex scalar that breaks 
the U(1)$'$ radiatively and a Dirac fermion which plays the role of sub-GeV dark matter candidate. The portal to the SM 
is provided by kinetic mixing of the $Z'$ gauge boson and, depending on the cosmological history, by thermal contact 
between the dark Higgs and the visible bath. The same supercooled MeV-scale phase transition that generates the $Z'$ 
and dark-matter masses also sources a SGWB in the PTA band. Ref.~\citep{Balan:2025SubGeVDM} identifies two benchmark
cosmologies: a ``coupled'' dark sector, in which the dark Higgs remains in equilibrium with the SM plasma until the phase
transition, and a ``secluded'' case, in which the dark sector decouples earlier and reheats separately. In both cases 
a global analysis has been performed including the PTA likelihood and dark-matter constraints, and strongly supercooled, 
slow FOPTs have been found at $T_{\rm reh}\sim\mathcal{O}(10~\mathrm{MeV})$ that reproduce the amplitude and slope 
of the NANOGrav signal when combined with a modest SMBHB component as demonstrated in Fig.~\ref{fig:Balan-PTA}. In this model,
the combined FOPT+SMBHB signal provides a substantially better fit to the PTA likelihood than a single SMBHB power law.

The qualitative picture in other types of conformal hidden sectors (see e.g.~Ref.~\citep{Li:2025HiddenThermal}) remains similar: a strongly supercooled transition at $T_*\sim{\rm MeV}$-${\rm GeV}$ produces 
a sharply peaked nano-Hz GW spectrum, with its amplitude and shape controlled by the degree of supercooling and 
the details of reheating in the dark sector.

More generic non-conformal dark Higgs sectors, in which a dark scalar acquires a vacuum expectation value through 
a renormalisable potential with explicit mass terms, can also feature strongly first-order transitions at MeV-GeV scales.
Typical examples involve a dark $U(1)$ or non-Abelian gauge group spontaneously broken by one or more complex scalars, 
possibly coupled to additional fermions that act as dark matter candidates~\citep{Nakai:2020oit,Bringmann:2020ura,
Addazi:2020DarkPT}. The resulting scalar potential can exhibit a tree-level barrier between phases or develop 
one through finite-temperature corrections, closely paralleling familiar electroweak extensions~\citep{Profumo:2007wc,Espinosa:2011ax,Caprini:2015zlo}.

In such non-conformal models, the transition strength and time-scale are controlled by the dark Higgs self-couplings, gauge
couplings and portal interactions with the SM. For moderate supercooling, one often finds $\alpha \sim 0.1$--$1$ and
$\beta/H_*\sim 10^2$--$10^3$, leading to relatively narrow GW peaks. However, by tuning the potential towards 
near-degeneracy of minima or by including additional scalar fields, much stronger supercooling is possible, with 
correspondingly larger GW amplitudes and smaller $\beta/H_*$~\citep{Schwaller:2015tja,Addazi:2020DarkPT}. 
These models have been proposed as explanations of the NANOGrav signal in terms of dark-sector FOPTs that 
do not rely on exact conformal symmetry, and they offer complementary phenomenology in dark-matter and collider
searches.
A further possibility is that the dark sector is not fully thermalised with the SM, so that the dark temperature differs 
from the photon temperature at the time of the transition. This can significantly modify the mapping between $T_*$ and 
the observed peak frequency and amplitude, and alters constraints from BBN and the CMB on extra radiation and entropy
injection~\citep{Cruz:2023HotNEDE,Bringmann:2020ura}. 

\subsubsection{Strongly-coupled confining dark sectors}
\label{sec:composite-dark}

Beyond weakly coupled dark Higgs sectors, strongly interacting dark sectors with their own confined gauge dynamics offer 
another compelling route to nano-Hz GWs. In such models a new non-Abelian gauge group confines at a scale $\Lambda_{\rm D}$,
giving rise to a spectrum of dark mesons and baryons that can act as composite dark matter candidates. The confinement
transition itself can be first order over a broad range of parameters, as suggested by lattice studies of QCD-like theories 
and their finite-temperature phase diagrams~\citep{Aoki:2006we,Borsanyi:2010bp}. If $\Lambda_{\rm D}$ lies in the MeV-GeV range, the associated confinement FOPT 
can generate a GW signal peaking in the PTA band.

Explicit realizations include dark QCD sectors with fermions in different representations, dark technicolour-like models 
and composite Higgs scenarios. In many cases the confinement transition is strongly first order, with a substantial change 
in the effective number of degrees of freedom, leading to sizeable $\alpha$ and relatively small
$\beta/H_*$~\citep{Schwaller:2015tja,Pasechnik:2024Composite}. The resulting GW spectra typically exhibit a single broad 
peak whose position is set by $\Lambda_{\rm D}$ and whose amplitude can approach the sensitivity of current PTAs and SKAO.
Depending on the matter content, the confined spectrum can include stable glueballs, dark baryons or other composites that 
act as dark matter, with relic abundances determined by confinement-scale annihilations, cannibalization or late
decays.

\begin{figure}[h!]
\centering
\includegraphics[width=0.6\textwidth]{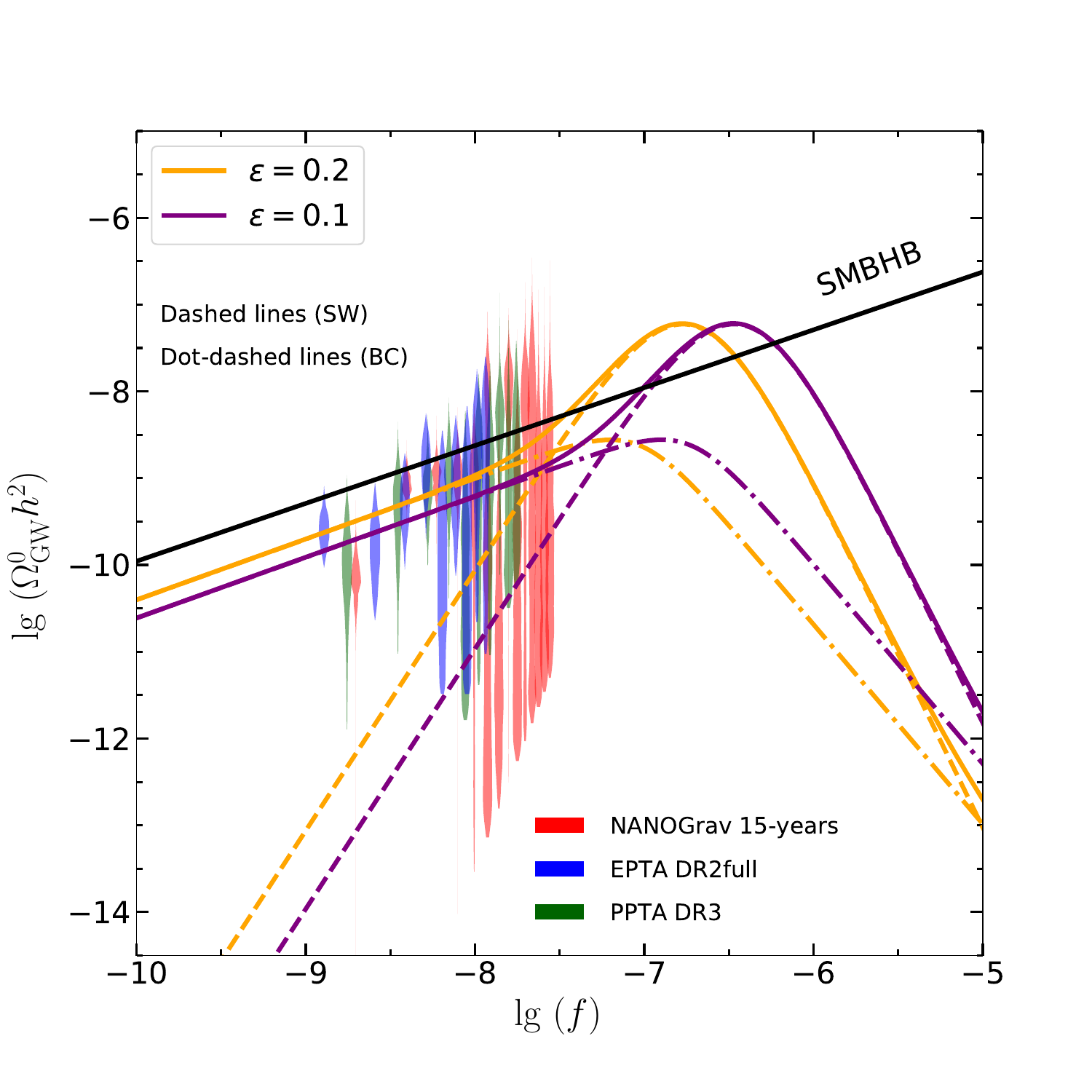}
\includegraphics[width=0.73\textwidth]{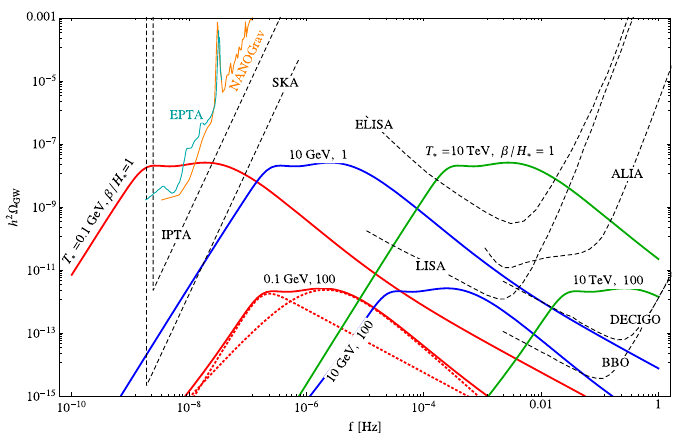}
\caption{GW signals from strongly coupled dark-sector phase transitions. 
    \emph{Top:} 
    Predicted stochastic GW background from a dark QCD phase transition in the mirror twin Higgs (MTH) framework, 
    adapted from Ref.~\citep{Zu:2024MirrorQCD}. The dash–dotted curves show the contribution from bubble collisions (BC) 
    and the dashed curves that from sound waves (SW); the solid curves give the total spectrum. Orange and purple lines
    correspond to different temperature ratios $\epsilon = T_{\rm twin}/T_{\rm SM} = 0.2,\,0.1$, respectively. 
    Coloured shaded regions indicate the tentative SGWB signals reconstructed by NANOGrav, EPTA and PPTA, while 
    the black line represents a fiducial SMBHB background. 
    \emph{Bottom:} 
    GW spectra $\Omega_{\rm GW}(f)h^2$ from a generic dark SU(N$_d$) confinement transition with different confinement 
    scales against sensitivity curves of major GW facilities, adapted from Ref.~\citep{Schwaller:2015tja}. 
    The three bands correspond to benchmark models with transition temperatures $T_* = 0.1~\mathrm{GeV}$ 
    (SIMP-like dark pions), $T_* = 10~\mathrm{GeV}$ (composite dark-matter and twin-Higgs–like scenarios) 
    and $T_* = 10~\mathrm{TeV}$, each shown for $\beta/H_* = 1$ and $100$, and bubble-wall velocity $v_w=1$.
  }
  \label{fig:dark-confinement-GW}
\end{figure}

Recent work has begun to combine lattice simulations, effective field theory and phenomenological modeling to quantify 
these signals more precisely. Studies of candidate hypercolour and hyper-fermion sectors relevant for composite Higgs 
and dark matter have explored the order of the confinement transition and its associated thermodynamics
\citep{Chen:2023QCDPT,Wang:2024StrongDS}. The resulting spectra 
illustrate how dark confinement transitions can source a nano-Hz background in the PTA/SKA band. 
Figure~\ref{fig:dark-confinement-GW} collects two representative examples. On the top, a mirror QCD phase transition 
in the mirror twin Higgs (MTH) framework produces a SGWB signal in the nHz range that can fit the NANOGrav, EPTA and PPTA 
data for suitably chosen values of the transition strength and duration~\citep{Zu:2024MirrorQCD}. On the bottom, a more 
general analysis of composite dark-sector models with an SU(N$_d$) confining gauge group~\citep{Schwaller:2015tja} shows 
how the peak frequency and amplitude of the GW signal scale with the confinement scale and $\beta/H_*$, and how GeV/sub-GeV
confinement transitions are naturally targeted by PTAs and future SKAO observations. 

These and other scenarios explored in the literature suggest that a variety of strongly coupled dark sectors can undergo 
first-order confinement transitions with parameters well suited to generate nano-Hz GWs, sometimes accompanied by 
additional relics such as long-lived bound states or primordial black holes (PBHs). Thus, confining dark sectors form 
a key benchmark class that directly links nano-Hz SGWB measurements to the microphysics of composite dark matter 
and non-perturbative dynamics beyond the SM.

\subsection{Primordial black holes from FOPTs}
\label{sec:pbh}

FOPTs are not only powerful sources of stochastic GWs, but can also seed the formation of PBHs. In strongly supercooled
transitions the Universe can briefly become vacuum dominated, the dynamics of bubble nucleation and collision can induce 
large inhomogeneities in the energy density, and regions of over-density may collapse into PBHs once they re-enter 
the horizon. The GW spectrum and the PBH population then become two complementary relics of the same transition, 
with their properties tightly correlated with the transition temperature and dynamics.

In a radiation-dominated Universe, the characteristic mass of PBHs formed around a temperature $T_{\rm form}$ is set 
by the mass contained in the Hubble volume at that time,
\begin{equation}
  M_{\rm PBH} \;\sim\; \gamma\, M_H(T_{\rm form})
  \;\simeq\;
  30\,M_\odot\,\gamma
  \left(\frac{g_*(T_{\rm form})}{10}\right)^{-1/2}
  \left(\frac{T_{\rm form}}{100~{\rm MeV}}\right)^{-2}\!,
  \label{eq:MPBH-scaling}
\end{equation}
where $\gamma\sim 0.1$ encodes the collapse efficiency and $M_H$ is the horizon mass. For MeV/GeV-scale FOPTs 
relevant to nano-Hz GWs this relation points to PBHs in the stellar to intermediate-mass range, which have been 
widely discussed as potential constituents of dark matter~\citep{Carr:2020PBHReview,Sasaki:2016jop}.

Several mechanisms have been proposed by which FOPTs generate the over-densities needed for PBH formation. 
In models with extreme supercooling, large bubbles nucleate in a nearly empty false vacuum, collide and leave 
behind strongly inhomogeneous reheated regions. The resulting fluctuations in the radiation fluid can reach 
the threshold for collapse, especially if the energy release is highly localized~\citep{Lewicki:2024SlowPT,Gouttenoire:2024PBH}.
In other scenarios the FOPT induces a sharp feature in the curvature power spectrum, for example through couplings 
between the order parameter and the inflaton or a spectator field, allowing local enhancements that later
collapse~\citep{Liu:2021svg,Gouttenoire:2023rxy,Gouttenoire:2023ftt}. In both cases the same parameters 
that control the GW spectrum -- in particular $\alpha$, $\beta/H_*$ and $T_*$ -- enter the PBH formation efficiency.

Recent work has explored in detail the interplay between PBH formation and nano-Hz GW signals from supercooled FOPTs. 
On the one hand, a very strong transition with $\alpha\gg 1$ and small $\beta/H_*$ favours PBH production but can 
also enhance the GW amplitude to the point where it is constrained by PTAs or CMB and BBN bounds on extra
radiation~\citep{Gouttenoire:2024PBH,Baker:2025PBHPT,Flores:2024Revisit}. On the other hand, the requirement that 
the transition completes and avoids trapping the Universe in a metastable phase restricts how extreme the supercooling 
and vacuum domination can be. Detailed analyses in conformal and non-conformal dark-sector models show that 
it is possible to simultaneously account for a fraction of the dark matter in PBHs and generate a nano-Hz GW background 
within reach of PTAs and SKAO, but only in relatively narrow regions of parameter
space~\citep{Liu:2021svg,Pasechnik:2025ConformalPBH}.

\begin{figure}[t]
\includegraphics[width=0.47\textwidth]{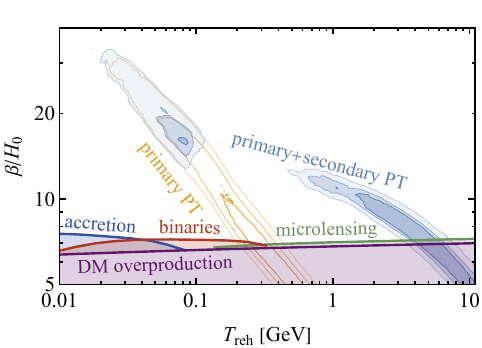}
\includegraphics[width=0.50\textwidth]{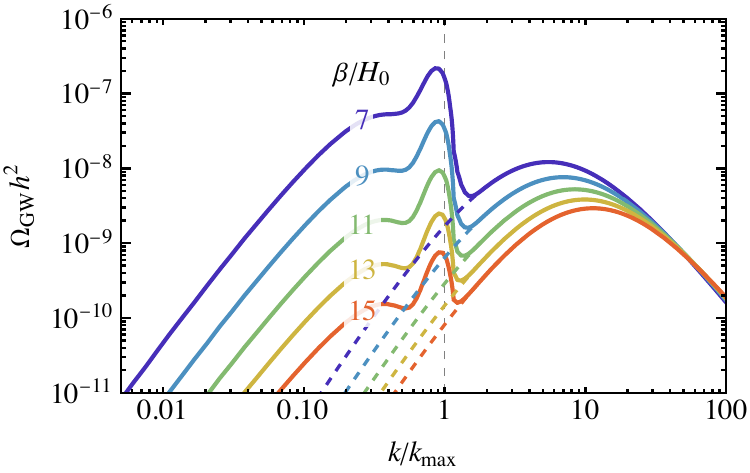}
\caption{
    GWs and PBHs from slow, strongly supercooled FOPTs, adapted from Ref.~\citep{Lewicki:2024SlowPT}. 
    \emph{Left:} 
    Fit of the GW spectrum to the NANOGrav 15-year data when both the primary (bubble-collision) and secondary 
    (scalar-induced) contributions from the transition are included (blue contours) compared to a fit using 
    only the primary spectrum (orange contours). The axes show the reheating temperature $T_{\rm reh}$ and 
    the inverse duration parameter $\beta/H_0$. Coloured regions at the bottom are excluded by constraints 
    on PBHs produced during the transition (evaporation, microlensing, accretion, binary mergers and 
    dark-matter overproduction).
    \emph{Right:} 
    Total present-day GW energy density $\Omega_{\rm GW} h^2$ as a function of wavenumber $k/k_{\rm max}$ for 
    several values of $\beta/H_0$. Solid curves show the sum of the primary and secondary components, while 
    dashed curves show the primary spectrum alone.
  }
  \label{fig:PBH-slowPT}
\end{figure}

The interplay between PBH formation and the GW spectrum in such slow, strongly supercooled transitions has been quantified 
in detail in Ref.~\citep{Lewicki:2024SlowPT}. In that work, the analysis was done for both the primary GW background from 
bubble collisions and fluid shells and the secondary, scalar-induced component sourced by the large curvature perturbations 
that also produce PBHs. The resulting signal consists of two characteristic peaks, whose relative importance depends 
mainly on $\beta/H_0$, while the same parameter space is tightly constrained by PBH abundance limits. 
Figure~\ref{fig:PBH-slowPT} illustrates the region favoured by the PTA data and the associated double-peaked GW spectra.

From the observational point of view, a PBH-FOPT connection implies characteristic correlations between the GW spectrum, 
the PBH mass function and other cosmological probes. For example, MeV-scale FOPTs that produce solar-mass PBHs generate 
GW spectra peaking squarely in the PTA band, while GeV-scale transitions associated with lighter PBHs peak at slightly 
higher frequencies. Constraints on PBHs from microlensing, CMB spectral distortions, large-scale structure and 
black-hole merger rates can therefore be combined with PTA/SKAO data to test the consistency of PBH-FOPT scenarios 
in a highly non-trivial way~\citep{Carr:2020PBHReview,Gouttenoire:2024PBH,Baker:2025PBHPT,Flores:2024Revisit}.

\subsection{Cosmic strings and other topological relics from symmetry-breaking FOPTs}
\label{sec:strings}

Spontaneous symmetry breaking in the early Universe generically leads to the formation of topological defects, 
whose nature is determined by the topology of the vacuum manifold~\citep{Kibble:1976sj,Vilenkin:2000jqa}. 
When the symmetry breaking is associated with a cosmological FOPT, the resulting defect networks can coexist with, 
and sometimes dominate over, the GW background from bubble dynamics and plasma motions. Cosmic strings -- line-like 
defects associated with the breaking of a $U(1)$ or more general gauge symmetry -- are particularly important 
in this context, as they produce a characteristic scale-invariant GW background that extends over many decades 
in frequency~\citep{Vilenkin:2000jqa,Sousa:2024CSGW}.

In gauge theories, cosmic strings form when the first homotopy group of the vacuum manifold is non-trivial. 
The Kibble mechanism implies that different Hubble patches choose uncorrelated vacua, leading to the formation 
of a network of long strings and loops. The network evolves towards a scaling regime in which its energy density 
remains a fixed fraction of the total, and energy is continuously shed into gravitational radiation via 
the oscillation and decay of loops~\citep{Kibble:1976sj,Vilenkin:2000jqa,Damour:2000wa,Damour:2001bk,Bohe:2011rk}. 
The resulting GW background has a nearly flat spectrum in $\Omega_{\rm GW}(f)$ over a wide range of frequencies, 
with small features associated with changes in the relativistic degrees of freedom and the loop distribution~\citep{BlancoPillado:2018ael}.

This spectral shape is quite distinct from the peaked spectra produced by individual FOPTs, but can nevertheless 
mimic a power-law SGWB in the relatively narrow frequency window currently probed by PTAs. Indeed, cosmic strings 
have emerged as one of the leading non-astrophysical interpretations of the NANOGrav signal, with viable fits obtained 
for string tensions $G\mu \sim 10^{-11}$-$10^{-7}$ depending on assumptions about the loop distribution and reconnection
probability~\citep{Ellis:2020ena,Blasi:2020mfx,Gouttenoire:2019kij,Ellis:2023tsl,Servant:2023UHF,Kitajima:2023cek}. 
Cosmic superstrings, which can arise in string-theoretic constructions and have reduced intercommutation probabilities, 
further enrich this phenomenology by supporting multiple string types and
junctions~\citep{Gouttenoire:2019kij,Ellis:2023tsl}.

Cosmic string networks can themselves be generated or affected by FOPTs. For example, if a global or gauged $U(1)$ symmetry 
is restored at high temperature and broken via a FOPT in a dark sector, the transition both sources a peaked GW component 
and seeds a string network that contributes a nearly scale-invariant
background~\citep{Buchmuller:2023TopDef,Fu:2023GlobalBL}. Hybrid scenarios combining FOPTs and strings 
have also been studied in grand-unified and left-right symmetric models, where cosmic strings and domain walls appear at
successive symmetry-breaking steps~\citep{Dunsky:2022FCOS,Buchmuller:2023TopDef,Tan:2025StringCosmo}. 
In such cases the total SGWB can be a superposition of a peaked component from bubble dynamics and a broad string-induced
plateau, with non-trivial implications for spectral fits to PTA
data~\citep{Borah:2023CSreview,Sousa:2024CSGW}.

\begin{figure}[h!]
\centering
\includegraphics[width=0.67\textwidth]{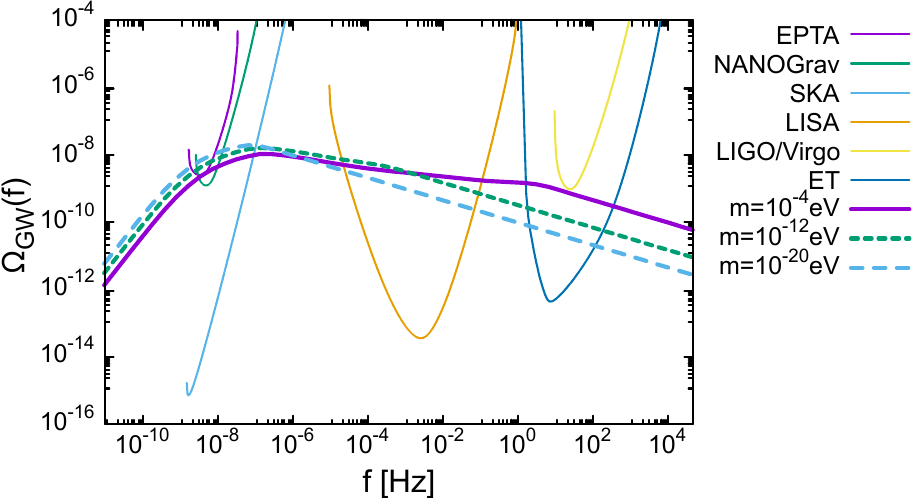}
\includegraphics[width=0.85\textwidth]{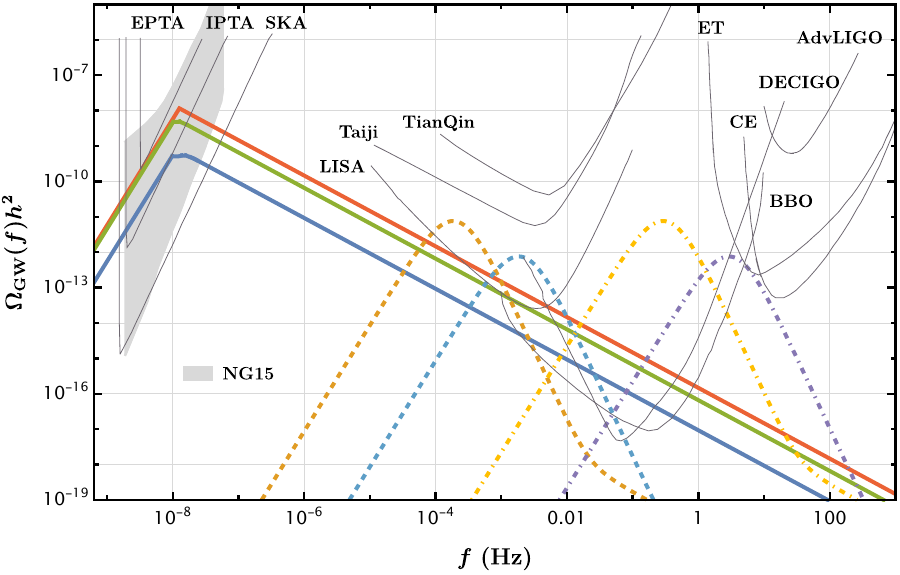}
\caption{
    GW spectra from FOPT-related topological relics, adapted from Ref.~\citep{Kitajima:2023cek} (top) and Ref.~\citep{Chen:2023QCDPT} (bottom). 
    \emph{Top:} 
    SGWB from a cosmic-string network formed by the breaking of a dark U(1) gauge symmetry, in the scenario where 
    cusps on string loops dominate the GW emission~\citep{Kitajima:2023cek}. 
    The curves show $\Omega_{\rm GW}(f)$ for a symmetry-breaking scale $v = 2\times 10^{12}\,\mathrm{GeV}$ and 
    dark-photon masses $m_{A'} = 10^{-4},\,10^{-12},\,10^{-20}\,\mathrm{eV}$.  
    Also shown are current PTA bounds (NANOGrav, EPTA) and projected sensitivities of SKA, LISA, LIGO/Virgo and the ET.
    \emph{Bottom:} 
    GW spectroscopy in QCD-collapsed domain-wall models~\citep{Chen:2023QCDPT}. 
    Solid lines show GW spectra from the annihilation of domain walls associated with a $Z_N$ symmetry anomalous 
    under QCD, for representative choices $(N,\sigma) = (2,1\times 10^{16}\,\mathrm{GeV}^3)$ (red), 
    $(4,3\times 10^{15}\,\mathrm{GeV}^3)$ (green) and $(6,1\times 10^{15}\,\mathrm{GeV}^3)$ (blue).
    Dashed lines denote additional contributions from a first-order QCD phase transition at $T_n^{\rm QCD}$ 
    with $(\alpha_{\rm GW},\beta_{\rm GW}/H) = (0.5,10^4)$ (orange) and $(0.5,10^5)$ (blue), while dot–dashed 
    lines correspond to a possible first-order discrete-symmetry transition at $T_{\rm form}$ with 
    the same parameter choices (yellow and purple). 
    The grey band indicates the range of spectra compatible with the NANOGrav signal, and coloured curves 
    show the sensitivities of current and future detectors including SKA, IPTA, EPTA, LISA, DECIGO and 
    ground-based interferometers.}
  \label{fig:CS-DW}
\end{figure}

A concrete realization of a FOPT-related topological source in the nano-Hz band is provided by cosmic strings associated 
with a broken dark U(1) gauge symmetry that also generates dark photon dark matter~\citep{Kitajima:2023cek}. In this scenario 
a complex scalar field $\Phi$ with potential $V(\Phi) = \frac{\lambda}{4} (|\Phi|^2 - v^2)^2$ spontaneously breaks the dark
U(1), giving mass to the dark photon, $m_{A'} = \sqrt{2}\,g\,v$, and forming a network of cosmic strings if the transition
occurs after inflation. The long-string network evolves towards a scaling regime and continuously sheds loops. Large loops
radiate GWs, while smaller loops efficiently emit dark photons, which can account for the observed dark-matter abundance 
for $m_{A'} \sim 10^{-6}$-$10^{-4}\,\mathrm{eV}$ and $v \sim 10^{12}\,\mathrm{GeV}$. The resulting stochastic background 
has a relatively hard spectrum around $f\sim 10^{-8}\,\mathrm{Hz}$, can match the PTA signal, and exhibits a characteristic 
high-frequency tail whose amplitude depends on the dark-photon mass and on whether GW emission is dominated by cusps 
or kinks on the loops. As suggested by Fig.~\ref{fig:CS-DW} (top), SKAO, together with LISA, LIGO/Virgo and ET, will be able 
to probe different parts of this multi-band spectrum and thereby test the consistency of the cosmic-string/dark-photon
interpretation of the nano-Hz excess.

Another illustrative example involves QCD-anomalous discrete symmetries and the associated domain walls~\citep{Chen:2023QCDPT}.
Here a discrete $Z_N$ symmetry is spontaneously broken at a high scale $f \sim 10^5$-$10^6\,\mathrm{GeV}$ by the vacuum
expectation value of a complex scalar $S$, leading to $N$ degenerate vacua and a network of domain walls separating them. 
The phase of $S$ couples anomalously to QCD through heavy coloured fermions, so that QCD instantons generate an effective
$\theta$-dependent potential at temperatures around the QCD scale and lift the degeneracy between the vacua. As the Universe
cools, this QCD-induced bias drives the collapse of the domain-wall network, generating a stochastic GW background whose peak
frequency and amplitude are controlled by the wall tension $\sigma \sim f^3$ and the bias scale. A distinctive feature of this
framework is that QCD itself may undergo a FOPT in domains with an effective $\theta$ angle close to $\pi$, which produces 
an additional GW component at somewhat higher frequencies. If the initial $Z_N$-breaking transition at $T_{\rm form}\sim f$ is
also first order, the model predicts a ``GW spectroscopy’’ with three FOPT-related contributions: domain-wall annihilation
around $T\sim 100\,\mathrm{MeV}$, a first-order QCD transition at slightly higher frequencies, and a high-frequency signal 
from the discrete-symmetry FOPT. As illustrated in Fig.~\ref{fig:CS-DW} (bottom), the combined spectrum spans more than ten
orders of magnitude in frequency, from the PTA/SKAO band up to ground-based interferometers, providing a powerful target 
for multi-band GW searches.

PTAs and SKAO play a crucial role in testing these scenarios. The near power-law spectrum suggested by current PTA data 
is naturally compatible with string-induced backgrounds, but future measurements of the spectral slope, possible breaks 
and anisotropies will help distinguish between pure string models, pure FOPTs and mixed
scenarios~\citep{Ellis:2023tsl}. Moreover, cosmic string tensions inferred from nano-Hz GWs can be 
related to symmetry-breaking scales in the underlying particle-physics models, sometimes directly probing seesaw 
scales, dark-matter properties or grand-unification physics~\citep{Buchmuller:2023TopDef,Fu:2023GlobalBL,Tan:2025StringCosmo}.

\section{SKAO prospects for probing cosmological FOPTs}
\label{sec:skao-prospects}

The SKAO will host the most sensitive radio interferometer ever built, with SKA-Mid and SKA-Low providing wide 
frequency coverage and exquisite sensitivity across a broad range of sky positions. In the context of GW astronomy, 
SKAO’s most important role is as a PTA, in which high-precision timing of a large sample of MSPs is used to detect 
the correlated timing deviations induced by GWs passing between the Earth and the pulsars~\citep{Detweiler:1979wn,Weltman:2020gw,Janssen:2015SKA,Stappers:2018rsta}. Compared to the 
current PTAs such as NANOGrav, EPTA/InPTA, PPTA and CPTA, SKAO will dramatically expand both the number of precisely 
timed MSPs and the timing precision achievable for each pulsar.

Forecasts suggest that SKAO will time hundreds of MSPs with sub-microsecond, and for a subset tens-of-nanosecond, precision 
over baselines extending well beyond a decade~\citep{Janssen:2015SKA,Weltman:2020gw,Stappers:2018rsta}. This will push 
the accessible GW amplitudes down by more than an order of magnitude compared to current PTAs, and will widen the effective
frequency window, especially at the lowest frequencies where long baselines are crucial. The increased sky coverage and the
inclusion of many southern-hemisphere pulsars will also improve the angular resolution of the background, enabling more
sensitive tests of isotropy and searches for anisotropic or non-Gaussian components~\citep{Gair:2013PTAGW}.

From the point of view of cosmological FOPTs, these improvements translate into a substantial gain in sensitivity to GW 
spectra with peak frequencies in the range $f\sim 10^{-9}$-$10^{-7}$~Hz and characteristic amplitudes 
$\Omega_{\rm GW} h^2 \sim 10^{-10}$-$10^{-8}$, depending on the spectral shape and the observing strategy. SKAO will 
therefore be able not only to confirm or refute the tentative SGWB signals already reported by existing PTAs, but also 
to probe large regions of parameter space in the theoretical models for SGWB sources that are currently unconstrained. 
Moreover, by improving the measurement of the spectrum, angular correlations and possible time evolution of the nano-Hz
background, SKAO will provide crucial information for distinguishing cosmological FOPTs from astrophysical and alternative
cosmological sources.

The previous section have outlined how a wide range of cosmological FOPTs and associated relics can generate 
nano-Hz GWs at levels already probed by current PTAs. We have also seen that many of these scenarios naturally 
inhabit the MeV-GeV temperature range, and that they often correlate the GW signal with other observables 
such as dark matter, neutrino, PBHs or topological defects. The benchmark scenarios discussed above must ultimately 
compete with a variety of other proposed explanations for the nano-Hz SGWB. Astrophysical populations of SMBHBs 
are expected to generate a background with a characteristic strain spectrum $h_c(f)\propto f^{-2/3}$, corresponding 
to $\Omega_{\rm GW}(f)\propto f^{2/3}$ in the frequency range where most binaries are in their inspiral phase~\citep{Sesana:2013SMBHB}. Cosmic strings, domain walls and PBHs generally produce broader, often nearly 
scale-invariant spectra~\citep{BlancoPillado:2018ael,Servant:2023UHF,Borah:2023CSreview,
Sousa:2024CSGW}, while PBH formation scenarios can involve mixtures of peaked and power-law components~\citep{Carr:2020PBHReview,Sasaki:2016jop,Liu:2021svg,Gouttenoire:2023rxy}. The central challenge 
for SKAO is therefore not only to detect a SGWB, but also to use spectral, angular and statistical information 
to discriminate between these possibilities and to test the consistency of any FOPT interpretation.

The most basic discriminant is the frequency dependence of the background. A single FOPT typically produces 
a broken power-law spectrum, rising as $f^3$ at low frequency, peaking around $f_{\rm peak}$ and decaying with 
a model-dependent power at high frequency~\citep{Caprini:2015zlo,Caprini:2019egz,Hindmarsh:2020hop,Athron:2023uxb}. 
SMBHB-induced backgrounds instead follow a nearly pure power law over the PTA band, unless environmental effects 
or binary eccentricity introduce breaks or turnovers~\citep{Sesana:2013SMBHB}. Cosmic strings yield an extended 
plateau in $\Omega_{\rm GW}(f)$, with mild features associated with changes in the effective number of relativistic 
degrees of freedom and the loop distribution~\citep{BlancoPillado:2018ael,Ellis:2020ena,
Blasi:2020mfx,Ellis:2023tsl}. With its wider frequency lever arm and higher signal-to-noise ratio, 
SKAO will be able to resolve subtle departures from simple power laws, detect or constrain spectral peaks 
and breaks, and thereby rule out large classes of models whose predicted shapes are incompatible with the data.

Angular correlations and statistical properties provide a second set of discriminants. In the simplest case 
of an isotropic Gaussian background dominated by tensor modes, the cross-correlation pattern between pulsar 
pairs is given by the HD curve. Anisotropic backgrounds, for example from discrete SMBHB populations or from 
cosmic strings with a small number of loops, can induce deviations from this pattern and lead to a non-trivial 
multipole structure in the sky map of the SGWB~\citep{Gair:2013PTAGW}. Cosmic 
strings may also generate a population of resolvable bursts from cusps or kinks, whose statistics differ from 
those of a pure Gaussian background~\citep{Damour:2000wa,Damour:2001bk,Bohe:2011rk,Servant:2023UHF}. Domain walls 
or other defect networks may leave complex non-Gaussian structures~\citep{Caprini:2009dy}. By virtue of its large pulsar sample and improved sky coverage, SKAO will significantly 
refine measurements of the angular correlation function and the higher-order statistics of the timing residuals, 
allowing tests of isotropy, Gaussianity and parity that are beyond the reach of current PTAs.

A third layer of discrimination comes from consistency with cosmological and astrophysical constraints. 
PBH-producing FOPTs must respect bounds on the PBH abundance from microlensing, CMB distortions, large-scale 
structure and black-hole merger rates~\citep{Carr:2020PBHReview,Gouttenoire:2024PBH,Baker:2025PBHPT,Flores:2024Revisit}. 
Dark-sector FOPTs must be compatible with bounds from laboratory searches for dark matter and dark forces, neutrino 
experiments and collider data~\citep{Ellis:2020awk,Bringmann:2020ura}.
Combining SKAO’s nano-Hz GW measurements with these complementary probes will allow one to test the internal 
consistency of proposed FOPT explanations and to distinguish them from alternative cosmological or astrophysical models.

The full power of the nano-Hz GW portal will only be realized in conjunction with other frequency bands and messengers. 
Space-based interferometers such as LISA and TianQin will be sensitive to mHz GWs from higher-scale FOPTs and from cosmic
strings, while ground-based detectors probe even higher frequencies~\citep{Caprini:2019egz,Servant:2023UHF}.
CMB, large-scale structure and BBN measurements constrain extra radiation, primordial magnetic fields and non-standard 
thermal histories. Laboratory searches for 
dark matter, neutrino properties and hidden sectors probe the same parameter spaces that control the
FOPTs. It is therefore natural to view SKAO as one component of 
a broader multi-band and multi-messenger programme. Within this programme, nano-Hz GW observations play a unique role 
by directly probing the dynamics of MeV/GeV-scale phase transitions, while other experiments and observations constrain 
the associated particle content, cosmic relics and cosmological histories. The combined data set will enable increasingly 
sharp tests of early-Universe physics and of the models that link cosmological FOPTs to the nano-Hz GW sky.

\section{Summary and outlook}
\label{sec:summary}

The nano-Hz GW window is rapidly evolving from a theoretical curiosity into an empirical probe of the early Universe. 
Pulsar timing arrays (PTAs) have now reported compelling evidence for a common-spectrum, low-frequency stochastic process 
with angular correlations consistent with an isotropic SGWB. If this signal is indeed of cosmological origin, it opens 
a new observational portal onto New Physics at MeV-GeV scales and onto cosmological histories that are otherwise extremely
difficult to access. The Square Kilometre Array Observatory (SKAO), with its unprecedented pulsar-timing capabilities, 
will be central to exploring this new regime.

We reviewed the current PTA landscape and introduced the nano-Hz GW portal. We summarized how PTAs measure correlated 
timing residuals across an array of millisecond pulsars, the role of the Hellings-Downs (HD) correlation in establishing 
a gravitational origin of the signal, and the present status of the NANOGrav, EPTA/InPTA, PPTA and CPTA results. We also
highlighted the range of possible nano-Hz sources, from astrophysical populations of supermassive black hole binaries 
(SMBHBs) to a variety of cosmological mechanisms, and emphasized the need for accurate theoretical modeling in order 
to interpret any future detection.

On this foundation, we further reviewed the physics of cosmological first-order phase transitions (FOPTs) as natural 
nano-Hz GW sources. We recalled how finite-temperature effective potentials can give rise to metastable phases separated 
by barriers, and how thermal tunneling, yielding bubble nucleation, growth and percolation, encodes the key macroscopic
parameters that control the GW signal, namely the transition strength $\alpha$, the inverse time-scale $\beta/H_*$, 
the bubble-wall velocity $v_w$ and the characteristic temperature $T_*$. We reviewed the modern understanding of 
GW production from bubble collisions, long-lived acoustic waves and magnetohydrodynamic turbulence, and discussed 
the associated theoretical systematics. We also summarized the state of the art in numerical tools and effective-theory
approaches used to map microscopic particle-physics models onto GW spectra.

Within this framework, we overviewed several benchmark classes of FOPTs that are especially relevant for the PTA/SKAO band.
Supercooled conformal dark sectors, in which radiative scale generation drives strong FOPTs at MeV-GeV scales, naturally 
produce sharply peaked nano-Hz GW spectra, often tied to neutrino-mass generation or dark matter production. More generic 
non-conformal dark Higgs sectors can also realize strong FOPTs with tunable $(\alpha,\beta/H_*,T_*)$, sometimes in dark 
sectors that are thermally decoupled from the SM. Strongly coupled confining dark sectors provide another compelling 
avenue, linking first-order confinement transitions at low scales to composite dark matter candidates and to nano-Hz GWs, 
with parameters informed by lattice studies of QCD-like theories. We showed that, in all these cases, the same 
non-perturbative or radiative dynamics that generate masses and couplings for dark states can simultaneously yield 
GW signals in the PTA band.

We then turned to additional relics of FOPTs, which both enrich and constrain these scenarios. Strongly supercooled 
transitions can seed primordial black holes (PBHs) in the stellar to intermediate-mass range, correlating the nano-Hz GW
spectrum with the PBH mass function and with constraints from abundance, microlensing, CMB distortions, large-scale structure 
and merger rates. Symmetry-breaking transitions can generate networks of cosmic strings, domain walls and other topological
defects, whose scaling evolution leads to broad, often nearly scale-invariant GW backgrounds extending across many decades 
in frequency. Finally, many of the dark-sector FOPTs considered here are directly motivated by neutrino physics, baryogenesis 
or dark-matter phenomenology, tying the nano-Hz GW signal to neutrino masses and mixings, lepton-number-violating processes 
and dark-sector dynamics.

Furthermore, we outlined the prospects for SKAO to interrogate this rich landscape. SKAO will time hundreds of millisecond
pulsars with unprecedented sensitivity and sky coverage, enlarging the accessible frequency range and lowering the detectable 
GW amplitudes well beyond current PTA capabilities. This will sharpen the measurement of the nano-Hz spectrum, its angular
correlations and possible time evolution, enabling subtle departures from simple power laws to be resolved. In combination 
with improved modeling, this will make it possible to distinguish peaked FOPT-induced spectra from the power-law backgrounds
expected from SMBHBs, from the broad plateaus produced by cosmic strings and from the signatures of PBH formation. At the same
time, consistency with CMB, BBN and large-scale structure constraints on extra radiation and primordial magnetic fields, 
and with laboratory searches for dark matter, dark forces and neutrino properties, will provide powerful cross-checks 
of any proposed FOPT explanation.

Looking ahead, progress on both the theoretical and observational sides will be essential. On the theory side, further
refinements of finite-temperature effective-field-theory methods, non-perturbative lattice simulations and hydrodynamic
modeling are needed to reduce the uncertainties in $(\alpha,\beta/H_*,v_w)$ and in the GW templates for sound waves and
turbulence. Improved global fits that incorporate these theoretical error bars, and that combine PTA/SKAO data with LISA
forecasts, CMB and large-scale structure constraints and particle-physics searches, will sharpen our ability to discriminate
between competing scenarios. On the observational side, SKAO’s long baselines, increased pulsar sample and improved timing
precision will gradually transform nano-Hz GW astrophysics from a low-significance detection regime into a precision science
domain, with measurements of anisotropy, non-Gaussianity and, eventually, polarization in the GW background.

If a nano-Hz SGWB of cosmological origin is confirmed and its properties measured with sufficient accuracy, the implications 
for fundamental physics will be profound. FOPTs at MeV-GeV scales would provide direct evidence for
new sectors beyond the SM, reveal the dynamical mechanisms by which masses and couplings are generated, and potentially shed
light on the origin of dark matter, the baryon asymmetry and neutrino masses. Even a null detection with SKAO would place
stringent limits on a broad class of FOPT scenarios, constraining the allowed parameter space for conformal and confining 
dark sectors and for PBH and defect-based models. In either case, nano-Hz GW observations with SKAO will play a pivotal role 
in mapping the thermal history of the Universe and in testing the deep connections between gravity, particle physics and
cosmology.


\bibliographystyle{abbrvnat-maxbibnames4}

\bibliography{chapter}

\end{document}